\documentclass[twocolumn]{svjour3}          % twocolumn
\smartqed  % flush right qed marks, e.g. at end of proof
\usepackage{graphicx}
\usepackage{amsmath}
\usepackage{fixltx2e}
\usepackage[sort]{natbib}
\usepackage{times}
\usepackage{hyperref}

\journalname{Cognitive Neurodynamics}

\begin{document}

\title{The phase response of the cortical slow oscillation %\thanks{Grants or other notes
}
%\subtitle{}


\author{Arne Weigenand \and
	Thomas Martinetz \and
        Jens Christian Claussen %etc.
}


\institute{ Arne Weigenand$^\dagger$ \at
	    University of Luebeck\\
	    Institute for Neuro- and Bioinformatics\\
	    Graduate School for Computing in Medicine and Life Sciences, University of Luebeck\\
	    D-23538 L\"ubeck \\
	    \email{weigenand@inb.uni-luebeck.de}\\
	    Phone: +49 451 500 5309\\
	    Fax: +49 451 500 5502%  \\
	    \and	
	    Thomas Martinetz \at
	    University of Luebeck \\
	    Institute for Neuro- and Bioinformatics\\
	    D-23538 L\"ubeck \\
	    \email{martinetz@inb.uni-luebeck.de} 
	    \and
	    Jens Christian Claussen \at
	    University of Luebeck \\
	    Institute for Neuro- and Bioinformatics\\
	    D-23538 L\"ubeck \\
	    \email{claussen@inb.uni-luebeck.de}\\
	    \\
	    $^\dagger$ corresponding author
}

\date{Cognitive Neurodynamics 6(4), 367-375 (2012) \\ {DOI: 10.1007/s11571-012-9207-z}}
%\date{Received: 01.08.2011 / Accepted: date}

\maketitle

%\mbox{} \hfill
\noindent
\fbox{
\href{http://www.springerlink.com/content/1871-4080/}{\bf Cognitive
Neurodynamics, Volume 6 (4), 367-375 (2012)} 
}%fbox
\\

\begin{abstract}
Cortical slow oscillations occur in the mammalian brain during deep sleep
and have been shown to contribute to memory consolidation, an effect that
can be enhanced by electrical stimulation.
As the precise underlying working mechanisms are not known
it is desired to develop and analyze computational models of slow
oscillations and to study the response to electrical stimuli.
In this paper we employ the conductance based model of
Compte \emph{et al.} [J Neurophysiol \textbf{89}, 2707]
to  study the effect of electrical stimulation.
The population response to electrical stimulation depends on the timing of the stimulus with respect to the
state of the slow oscillation.
First, we reproduce the experimental results of
electrical stimulation in ferret brain slices by
Shu \emph{et al.} [Nature \textbf{423}, 288]
from the conductance based model.
We then numerically obtain the phase response curve for the conductance based network 
model to quantify the network's response to weak stimuli. 
Our results agree with experiments \emph{in vivo} and \emph{in vitro} 
that show that sensitivity to stimulation is weaker in the up than in the down state.
However, we also find that within the up state stimulation leads to a shortening of the up state, or phase
advance, whereas during the up-down transition a prolongation of up states
is possible, resulting in a phase delay.
Finally, we compute the phase response curve
for the simple mean-field model by 
Ngo \emph{et al.} [Europhys Lett \textbf{89}, 68002]
and find that the qualitative shape of the PRC is preserved, 
despite its different mechanism for the generation of slow oscillations.
 
\keywords{sleep \and cortex \and phase response \and slow oscillation \and synchronization}
\end{abstract}

\section{Introduction}
\label{intro}
%\begin{itemize}
% \item 	
During the deep sleep stages S3/S4 of mammalian sleep 
	the electroencephalogram (EEG) exhibits large amplitude
	oscillations at frequencies of 1Hz and below \citep{contreras_cellular_1995}.
	% oder anderes Zitat? 2x Massimini ist etwas viel. Steriade?
	These so-called slow oscillations are a phenomenon
	with a much slower time scale than that of a single spiking neuron and reflect 
	the alternation of periods of activity and silence of large neuronal populations.
	Cortical slow waves not only manifest an interesting dynamical phenomenon
	on its own,
	but also have been shown to significantly contribute to memory consolidation 
	in humans and other mammals
\citep{diekelmann_memory_2010,marshall_boosting_2006,stickgold_sleep-dependent_2005}.
	A consequent and appealing approach is therefore to
	enhance sleep slow waves by stimulation techniques
	with the goal of enhancing the consolidating effect on memories
	\citep{marshall_transcranial_2004,massimini_triggering_2007}.
	Therefore a more detailed understanding of the underlying dynamical
	mechanisms is desired to further develop stimulation techniques.

Networks of neurons often exhibit collective oscillations \citep{brunel_dynamics_2000,Gray_Konig_Engel_Singer_1989,jirsa2008},
during which single neurons spike irregularly \citep{hajos2004}. 
The collective dynamics are periodic though and one can treat the network as one large oscillator~\citep{akam2012,grannan_stimulus-dependent_1993}.
The cortical slow oscillation shows high temporal regularity in ferret brain slices
and in rat auditory cortex under deep anesthesia \citep{deco_effective_2009,sanchez-vives_cellular_2000,mattia2012} 
and can thus be characterized by a phase response curve (PRC).

In this paper we obtain, based on computational models, 
predictions for the PRC of the cortical slow oscillation for a wide range of stimulus strengths.
The PRC is a map that describes how an oscillating system responds to perturbations \citep{granada_chapter_2009} and can easily be measured experimentally. 
Phase models have a long tradition and were successfully applied to study the interaction of coupled oscillators
\citep{Tass_1999,kuramoto_84,winfree_2001}. % bitte pruefen...!
More recently phase response curves were used to characterize synchronization between cortex and thalamus 
during epileptic sei-zures \citep{perez_velazquez_phase_2007} and dentate gyrus �- CA3 coupling in the hippocampus \citep{akam2012}.

Knowing the PRC one has a valuable tool to analyse the influence of external stimulation, e.g. electric, magnetic or sensory stimulation, 
on cortical sleep rhythms and also to investigate 
the interaction of the sleeping cortex with other brain structures, like hippocampus and
thalamus. These interactions are assumed to be of substantial relevance for memory
consolidation and transfer of memories between brain regions~\citep{peyrache_replay_2009}.
%This interaction is thought to underly memory consolidation.

Up and down states that comprise the slow oscillation during mammalian sleep seem to be a robust dynamical
phenomenon across species and also across cortical brain regions~\citep{maclean_internal_2005,amzica_electrophysiological_1998,sanchez-vives_rhythmic_2008}.
Therefore one would conjecture that models for slow oscillations as well as
models for their stimulation should not crucially depend on model details --
albeit one has to specify a working model and its parameters for 
computational studies.
 
This paper is organized as follows. In section~\ref{sec:delay} we demonstrate that the network model introduced by 
\\Compte \emph{et al.} \citep{compte_cellular_2003} is
capable of reproducing the experimental results of Shu \emph{et al.} \citep{shu_turning_2003}. 
Second, we build on this result and argue that this model is a suitable candidate to predict the response to weaker stimuli.
We present phase response and phase transition curves for Type 1 (weak) and Type 0 (strong) resetting as well as for 
intermediate stimulus intensities that serve as predictions for experiments.
Third, we obtain the infinitesimal PRC from the mean-field model by Ngo \emph{et al.} \citep{ngo_triggering_2010}, a minimal model for up-down state dynamics.
We find that the network model and the mean-field model yield qualitatively similar results.

\section{Network model reproduces characteristic delay of up-down transition upon stimulation}
\label{sec:delay}
\begin{figure}[t]
\centering
\includegraphics[width = 76mm]{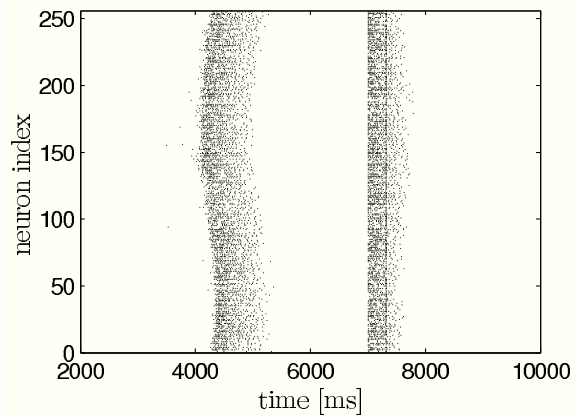}
\caption{Response of neural network to two consecutive strong stimuli ($ISI=310\mathrm{ms}$, $I_\mathrm s=1\mathrm{\mu A}$) as in \citep{shu_turning_2003}. 
The first stimulus causes an immediate transition from the down to the up state. 
The following second stimulus (straight line within second up state) determines the remaining time the system spends in the up state. 
It causes a massive influx of calcium which in turn activates the inhibiting $I_\mathrm{KCa}$ (not shown) that then leads to the termination of the up state. 
Only pyramidal neurons are shown. The stimuli are applied to each neuron in the network.}
\label{fig:Figure1}       % Give a unique label
\setcounter {figure}{1}
\end{figure}

\begin{figure}[t]
\centering
%  \includegraphics[width = 76mm]{./images/Shu.pdf}
\includegraphics[width = 76mm]{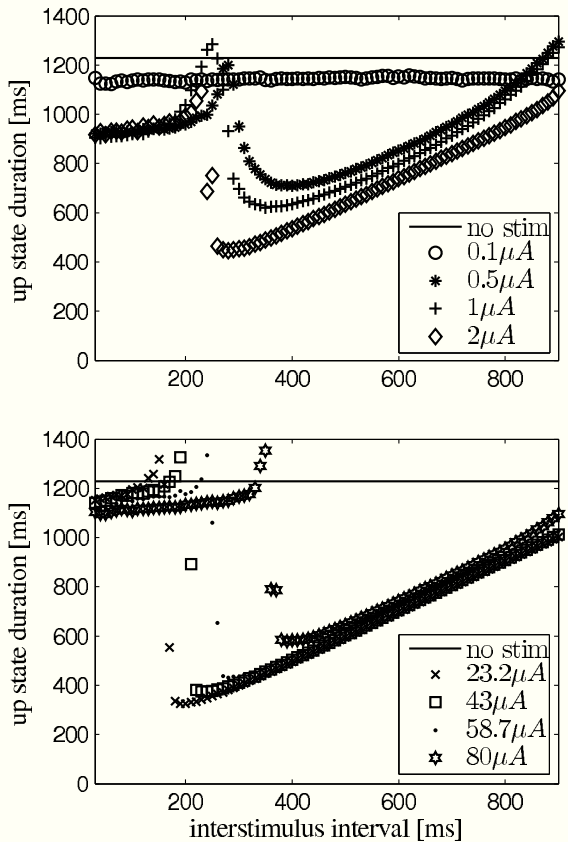}
\caption{Qualitative reproduction of the experimental results reported by Shu \emph{et al.} \citep{shu_turning_2003} with the 
network model. Data points are the average of 5 trials. Two depolarizing stimuli, separated by the interstimulus interval, where applied, see figure \ref{fig:Figure1}.
The peaks just before the transition to shorter up state durations that are visible in every curve are an artifact stemming 
from a heterogenous network response like the one shown in figure \ref{fig:Figure7}.
(top) Weak stimuli, e.g. $I_\mathrm s = 0.1 \mathrm{\mu A}$, that already cause strong resetting only reduce the up state duration, independent of phase.
Increasing the stimulus strength reduces the up state more the more the two stimuli are apart, until the second stimulus directly terminates an up state.
For certain stimulus strengths the second stimulus ends an up state immediately for almost all interstimulus intervals.
(bottom) In our simulations it was possible to evoke up state like network behavior also with very high stimulus strengths. 
This was different from mere after spiking. The higher the stimulus strength was the larger the interstimulus interval had to be in order to reduce up state durations.
This reversed tendency is not covered by \citep{shu_turning_2003} and remains to be tested experimentally.}
\label{fig:Figure2}       % Give a unique label
\setcounter {figure} {2}
\end{figure}

In this section we show that the network model introduced by Compte \emph{et al.}~\citep{compte_cellular_2003} is capable of qualitatively reproducing the experiment of Shu \emph{et al.} 
in \citep{shu_turning_2003}. 
Shu and colleagues showed that cortical activity can be switched on and off externally with excitatory stimuli. 
In their experiment two short current pulses of same polarity where applied to ferret brain slices exhibiting spontaneous slow oscillations. 
% The first pulse would cause a transition from down to up state. 
The second pulse was applied during the evoked up state and would lead to a termination of the up state after a certain delay.
That delay was consistent across trials and depended strongly on the stimulus amplitude and the actual interstimulus interval. 

The network model is conductance based and exhibits up-down state dynamics as were observed in ferret brain slices \emph{in vitro} \citep{sanchez-vives_cellular_2000}. 
The model proved its usefulness in recent studies \citep{sanchez-vives_rhythmic_2008,froehlich_endogenous_2010,sanchez-vives_inhibitory_2010}.
All details of the model can be found in the original paper \citep{compte_cellular_2003}. 
We restate the full equations in appendix \ref{sec:network_model}. 
In the following we only want to state some of its main features. 
The system contains $80\%$ regular spiking pyramidal neurons and $20\%$ fast spiking interneurons.
The pyramidal neurons possess two compartments and show spike frequency adaptation when seeing a constant injected current.
Pyramidal neurons are all excitatory and connect via AMPA and NMDA type synapses.
Inhibitory connections are only formed via $\mathrm{GABA_A}$ synapses.
The transition from the down to the up state is caused by spontaneously firing pyramidal neurons and recurrent excitation.
Importantly, the model does not require noise to switch between up and down states and exhibits self sustained activity without external drive.
The mechanism for the termination of up states is the activity dependent build up of inhibitory currents during the up state.
This occurs via a sodium dependent potassium channel whose activation increases with each spike.
The original model uses 1280 neurons in total. However, one can reduce the size of the system without changing the overall dynamics, if one also scales down 
the range of the synaptic connections accordingly. We compared the behavior of the system for different sizes and found no significant differences.
We therefore chose to work with a system size of only 320 neurons, because of the large number of simulations necessary for the results presented in this paper.

The network is stimulated two times with depolarizing current pulses of same polarity, intensity and duration. 
The pulses are applied to all neurons in the network at the same time. 
The pulse duration is 10 ms. The first stimulus is applied during the hyperpolarization phase inbetween two otherwise self�-generated up states. 
We implicitly assume that the external stimulation with electric shocks translates into a
transmembrane current that equally effects pyramidal neurons and interneurons. We also point out that stimulating all neurons
is in contrast to the experiment, where the stimulation was applied locally.
The protocol is illustrated in the raster plot (model data) in figure~\ref{fig:Figure1}. 
We applied the above stimulation protocol to the network model and yield a similar dependence of up state duration on stimulus amplitude and interstimulus interval. 
This is depicted in figure~\ref{fig:Figure2}. For comparison please see \citep{shu_turning_2003}. 

The protocol for obtaining a PRC is very similar to paired pulse stimulation.
Hence, if a model reproduces the response to a paired stimulus protocol it is likely that one can obtain the biologically realistic PRC from it. 
Our simulations show that the experimental results obtained by \citep{shu_turning_2003} are in the strong resetting regime. 

\section{The slow oscillation's PRC as indicated by network model and mean-field model}
We now present PRCs of the network model introduced above for weak resetting (infinitesimal PRC), strong resetting and intermediate stimulus intensities.  	
We compare the infinitesimal PRC of the network model with the infinitesimal PRC of the mean-field model (figure~\ref{fig:Figure5}) 
introduced by Ngo \emph{et al.} \citep{ngo_triggering_2010}. 
As in the network model the mechanism for terminating up states is the activity dependent build up 
of an inhibiting current. This is in contrast to rate models of the slow oscillation that are based on fluctuation-driven transitions between two stable fixed points
 \citep{deco_effective_2009,ermentrout_mathematical_2010,mejias_irregular_2010}.
Although the two models we used are of a different class and complexity they lead to PRCs with similar features.
% First, we define ensemble phase and perturbation of the neuron network and describe the numerical procedure to obtain the PRC from the network model.

\subsection{Phase response of network model}
\begin{figure}[t]
\centering
 % \includegraphics[width = 76mm]{./images/PhaseDef.pdf}
\includegraphics[width = 76mm]{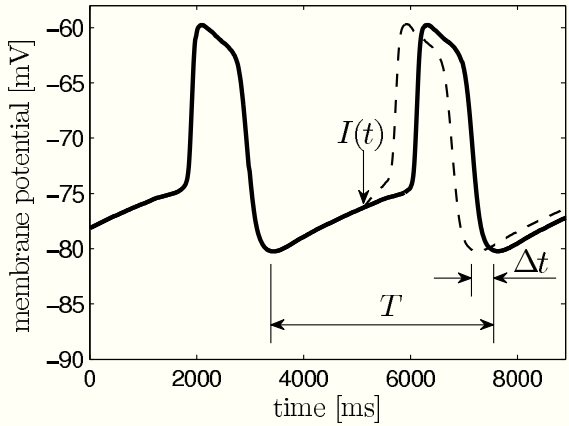}
\caption{Definition of phase resetting in network model and mean-field model. 
The solid line is the membrane potential trace produced by the network model averaged over all pyramidal neurons and smoothed subsequently. 
The perturbation $I(t)$ causes a phase reset that can delay or advance the oscillation (dashed line). 
We defined phases 0 and 1 to be the beginning of a down state/end of an up state. 
The phase reset is $\Delta \theta = \frac{\Delta t}{T}$.}
\label{fig:Figure3}       % Give a unique label
\end{figure}

\setcounter {figure} {3}
\begin{figure}[h]
\centering
  %\includegraphics[width=76mm]{./images/PRCvsI.pdf}
\includegraphics[width = 76mm]{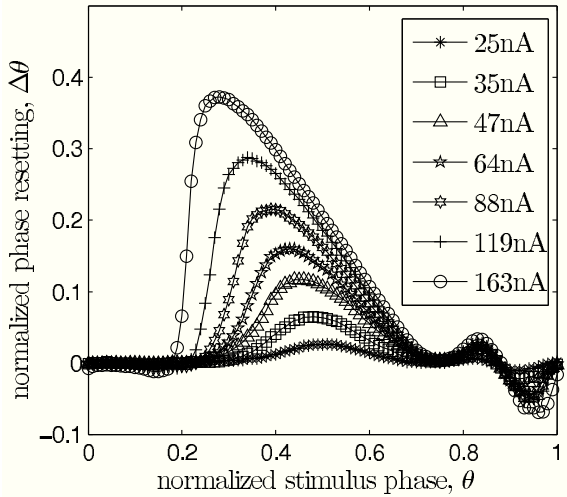}
\caption{Dependence of the network model's PRC on stimulus strength $I_\mathrm s$. The PRC tilts to the left as the stimulus strength increases.
Note that the phase resetting is only normalized to the oscillation period and not to $I_\mathrm s$.}
\label{fig:Figure4}       % Give a unique label
\end{figure}

\setcounter {figure} {4}
\begin{figure*}[h]
\includegraphics[width = 1\textwidth,height=7cm]{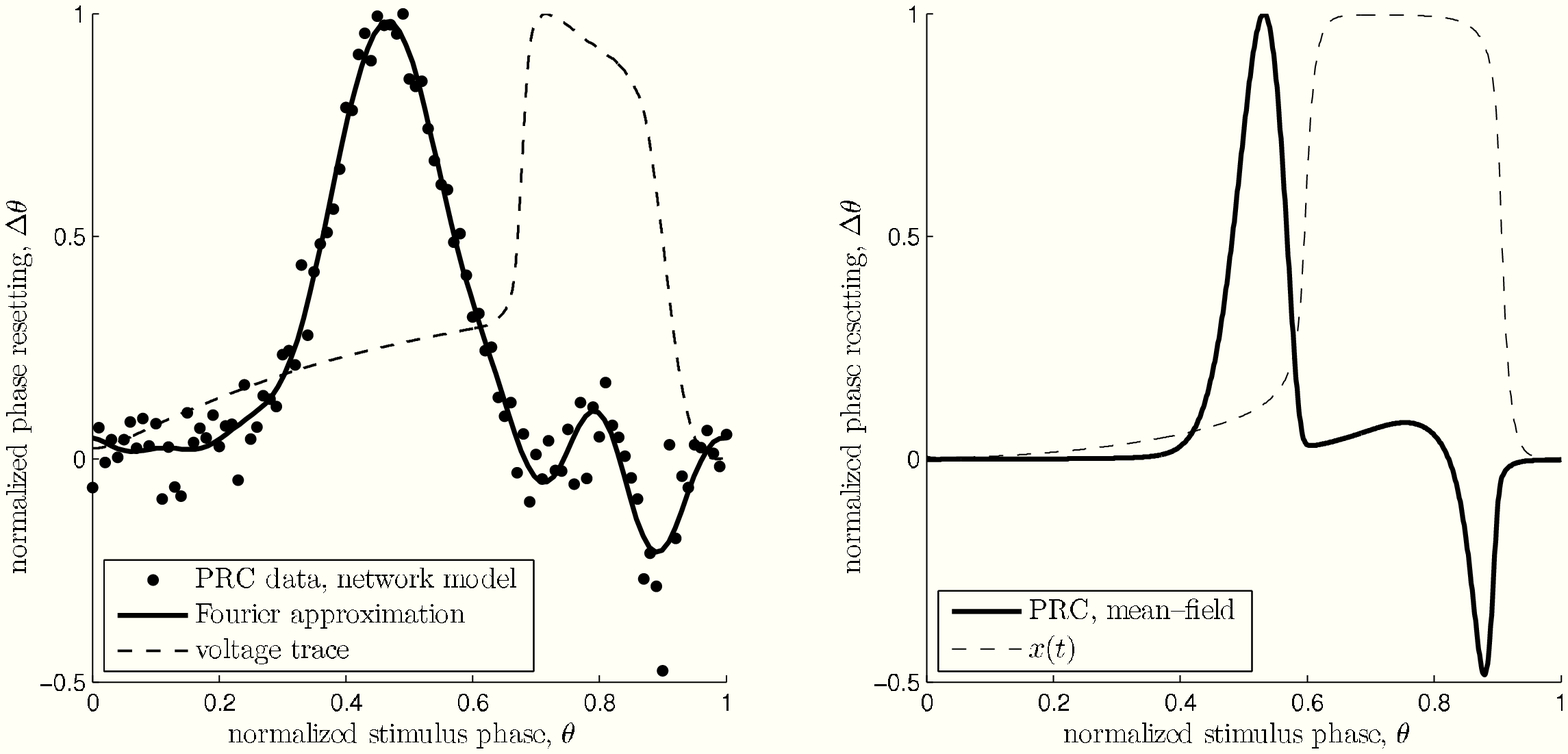}
\caption{Comparison of the two estimates of the slow oscillation's infinitesimal PRC. 
The dashed curves show the phase of the unperturbed oscillation in both plots and are in arbitrary units. 
Left: PRC of the network model for $I=19$nA. Black dots are from direct perturbation of the network at the respective phase $\theta$. 
The solid curve is a Fourier approximation of the data points of order 7.
The voltage trace was obtained by averaging one oscillation period over all pyramidal neurons and subsequent smoothing. 
Right: PRC of mean field model with $d_f=0.17$, $d_b=0.98$, $C=0.6$, $\sigma=0.05$, $\lambda_\nu=0.96$, $\lambda_\mu=0.9$, $g=0.1$,$h=0.2$.
The parameters where chosen to closely match the PRC of the network model. The model has a similar qualitative behavior over a wide range of parameters.
In both models stimulation is ineffective right after an up state. It has the largest impact at the end of the down state right before the transition to the up state. 
Within the up state, stimulation initially leads to a phase advance, i.e. a reduced up state duration. 
During the following up-down transition a phase delay is possible resulting in a prolonged up state.}
%\includegraphics[width = 76mm]{./images/Figure5.eps}
\label{fig:Figure5}       % Give a unique label
\end{figure*}
A phase response curve quantifies the response of a periodically oscillating system to a perturbing stimulus at a given phase.
We define the phase variable $\Theta$ as $\Theta = 2\pi t / T$, 
where $t$ denotes the elapsed time from the previous down state onset and 
$T$ is the period of the network oscillation \citep{Yasuhiro2007}. 
This is illustrated in figure~\ref{fig:Figure3}.
Onsets of up and down states were determined from the voltage trace of single neurons with the MAUDS algorithm \citep{seamari_robust_2007}.	
We define the ensemble phase of the network as the average phase of the individual neurons with respect to their down state onset.
The phase reset $\Delta \Theta$ is the phase difference between the perturbed and unperturbed neuron,
\begin{equation}
 \Delta \Theta = \overline \Theta - \Theta=\frac{\Delta t}{T},
\label{eq:1}
\end{equation}
where $\overline \Theta$ is the new phase immediately after the perturbation 
and $\Theta$ is the phase at which the stimulus was applied. 
Variables $\Delta t$ and $T$ are as in figure~\ref{fig:Figure3}.
The new phase is calculated from the simulation data via
\begin{equation}
 \overline \Theta = 1- \frac{t_\mathrm d - t_\mathrm{s}}{T}, 
\end{equation}
with $T$ being the oscillation period, $t_\mathrm d$ the beginning of the down state following the perturbation and $t_\mathrm{s}$
the time when the perturbation is applied. 
The old phase $\Theta$ is $(t_\mathrm{s}-t_2)/T$, where $t_2$ is the beginning of the down state before the perturbing stimulus.

The PRC can be determined using conductance\\ changes or current pulses as perturbation. 
It has been shown that both approaches are equivalent \citep{achuthan_synaptic_2010}.
We chose the latter option as it depends only on the intrinsic properties of a neuron.
To obtain the PRC we calculated $\Theta$ and $\overline{\Theta}$ of each pyramidal neuron for 50 different stimulus times. 
The perturbation is applied to all neurons at the same time but each neuron is in a slightly different phase with respect to the transition to its down state.
We used nearest neighbor interpolation and transformed the data points $(\Theta,\overline{\Theta})$ 
to an equidistant grid $\theta$ with step size 1/50 to facilitate averaging. 
Finally, the ensemble phase is determined using the circular mean $\overline{\theta}$ of the individual phases of pyramidal neurons in the network,
\begin{equation}
  \overline{\theta} = \mathrm{arg} \left( \frac{1}{N} \sum _{k=1}^N \mathrm e^{i 2\pi \overline{\Theta}_k}\right)
\end{equation}
and the phase reset $\Delta \theta$ is analog to \ref{eq:1}.
The infinitesimal PRC of the network model is depicted in figure~\ref{fig:Figure5} (left). It is renormalized to 1 for comparison with the mean-field model.
For stimulus amplitudes up to 19nA it scales linearly with stimulus amplitude.
Figure~\ref{fig:Figure4} shows the PRC's dependence on intermediate stimulus intensities.
For intensities between 19nA and 400nA the PRC is still qualitatively similar to the infinitesimal PRC, 
but does not scale linearly with stimulus intensities anymore.
\setcounter {figure} {5}
\begin{figure}[t]
\centering	
\includegraphics[width = 76mm]{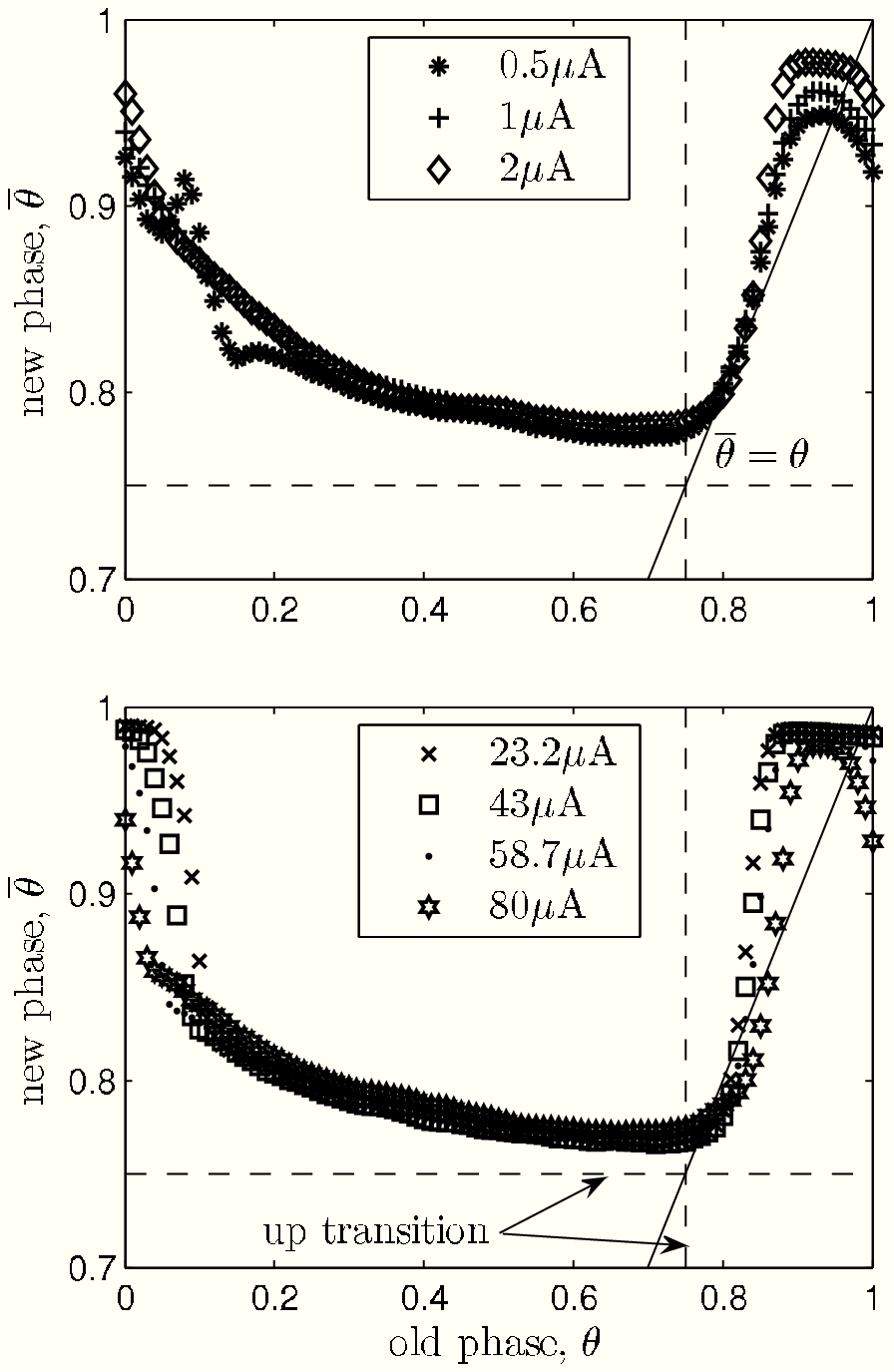}
\caption{Phase transition curves (PTCs) of the network model for Type 0 (strong) resetting. The solid line marks the condition $\overline{\theta}=\theta$, e.g. 
slope 1. The shortening of an up state that results from a stimulation at $\theta=[0.1\ldots 0.75]$ is almost independent of the stimulus intensity, 
as indicated by the overlapping curves in that range. Significant differences are apparent at the transition from up to down state and down to up state, respectively.
(top) The PTCs mostly stay above $\overline \theta=\theta$, indicating that in this intensity range up state durations can only be decreased. 
(bottom) The model predicts that there is a refractory period only for mediumly strong stimuli ($I_\mathrm s =[23.2,43,58.7]\cdot \mathrm{\mu A}$), 
as the phase transition curve is close to $\overline\theta =1$. 
Also, the slopes near the state transitions are steeper for strong stimuli. 
Hence it is more likely for very strong stimuli to have the desynchronizing effect shown in Figure \ref{fig:Figure7}.}
\label{fig:Figure6}       % Give a unique label
\end{figure}

% ---------------------------------------------------------
% We focus on the collective PRC of the excitatory pyramidal neurons because long distance connections are mostly exitatory.
\subsection{Phase reduction of mean-field model}
Ngo \emph{et al.} recently introduced a minimal model for the generation of cortical up and down states.
The original model of Ngo et al. is a time-discrete map. The full model, reformulated as system of differential equations, is
\begin{subequations}
\begin{align}
  \frac{\mathrm{d}x}{\mathrm{d}t}   &= \left(1 + \mathrm e^{-\beta(C x - d_f - \vartheta)} \right)^{-1}-x \\
  \frac{\mathrm{d}\mu}{\mathrm{d}t} &= \lambda_{\mu}\mu + g x-\mu \\
  \frac{\mathrm{d}\vartheta}{\mathrm{d}t} &= \lambda_{\vartheta}\vartheta + h\left(1+\mathrm e^{-\beta(\mu - d_b)}\right)^{-1} -\vartheta.
\end{align}
\end{subequations}
The variable x ranges between 0 and 1 and describes to what extent the population is active. 
$\mu$ is an activity dependent variable that increases when $x$ is active and could be interpreted as calcium current.
$\vartheta$ has an inhibiting effect on $x$ and is triggered by $\mu$. It could be interpreted as calcium dependent potassium current.
$\beta$ describes the noise level of the population, $C$ stands for the coupling strength,
$d_\mathrm f$ is a constant firing threshold 
and $\lambda_{\mu}$ and $\lambda_{\vartheta}$ are recovery rates of $\mu$ and $\vartheta$ respectively.
We then used the software \emph{XPPAUT} to numerically obtain the PRC \citep{ermentrout_XPP}. The result is shown in figure~\ref{fig:Figure5} (right). 
We chose the parameters of the mean-field  model to closely match phases of up and down states and PRC of the network model.
According to this model perturbations have the largest influence in a relatively short time window right before the transition to the up state 
and lead to a phase advance, i.e. a shortening of the down state. At the beginning of an up state perturbations also lead to a phase 
advance and a shortening of the up state, however only to a comparatively small extent. 
Perturbations toward the end of an up state have a larger impact, leading to a phase delay and hence can prolong the up state.

\setcounter {figure} {6}
\begin{figure}[htp]
\centering
\includegraphics[width = 76mm]{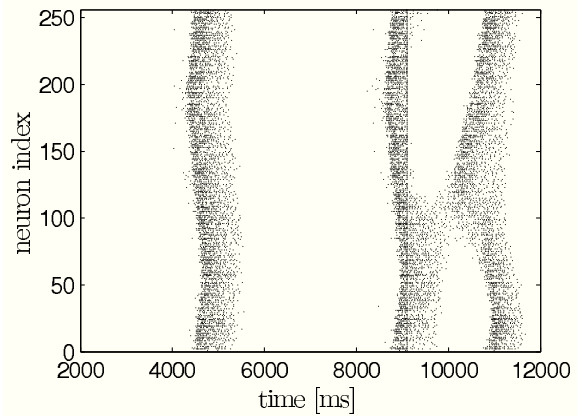}
\caption{Disrupting effect of a strong stimulus ($I_\mathrm s = 6.7\mu A$, $\theta \approx 0.85$) applied at phases with rapidly changing slope of the PRC for strong resetting, depicted in Figure~\ref{fig:Figure6}. 
As individual neurons never have identical phases when being in a collective up state it is possible to terminate the up state in one part of the network while at the same time extending it in another part, 
thus resulting in an effective desynchronization of the 1D system.}
\label{fig:Figure7}       % Give a unique label
\end{figure}

\section{Discussion}
In this paper we obtained a testable prediction for the PRC of the neocortex during deep anesthesia and for
slices of cortex tissue exhibiting up and down states. In the weak resetting regime we found type II PRCs with similar features for two different models 
that reproduce many aspects of up and down states in slices. The obtained PRCs show maximal responsiveness close to the transition to the up state.
This is in agreement with evoked potential studies \citep{massimini_2002} in humans and animals \citep{maclean_internal_2005,petersen2003}. 
In the strong resetting regime both models also conform to the experimental results by Shu et al. \citep{shu_turning_2003}.
Our results strictly apply only to ferret brain slices, as both investigated models build on observations from those preparations. 
However, considering the universality of sleep and related phenomena like spindles and hippocampal ripples across mammals our results should, 
at least qualitatively, translate to other species as well.

During natural deep sleep cortical slow oscillations are less regular than observed under certain kinds of anesthesia and in slice preparations. 
The reason for this is largely unclear.
Theoretical investigations assuming noise as driving force for the switching between up and down states 
predict a power law distribution \citep{mejias_irregular_2010} of the residence times in up and down states, but also 
showed that a purely fluctuation driven transition between up and down states is not sufficient 
to account for the statistics of residence times \citep{deco_effective_2009}.  
Rather, the probability density function obtained from experimental data is unimodal and centered on a 
preferred frequency not close to zero \citep{deco_effective_2009}.
The global dynamics of the conductance based network are that of a relaxation oscillator. 
The slow potassium currents in the model lead to a gradual build up of inhibition during the action of fast spiking currents in the up state 
and terminate it subsequently.
This inhibition then relaxes during the down state.
Although phase response theory fails in predicting the rapid synchronization behavior of relaxation oscillators~\citep{somers_waves_1995} 
it is appropriate for relaxation oscillators if coupling is weak and the oscillator is not close to the relaxation limit \citep{izhikevich_phase_2000,varkonyi98}.
The observation that the cortical slow oscillation propagates as a travelling wave~\citep{massimini_sleep_2004} supports this notion.

Phase response theory allows for accurate prediction of phase locking between oscillators and can be useful to analyze interactions between brain regions 
\citep{levnaji_phase_2010,ko_phase-response_2009,kori_collective-phase_2009,perez_velazquez_phase_2007}, especially their phase coherence \citep{akam2012}. 
During mammalian deep sleep hippocampal sharp wave ripple complexes and thalamic spindles tend to be phase-locked to the neocortical slow oscillation
\citep{molle_grouping_2002,clemens_temporal_2007,mayer_corticothalamic_2007} and parahippocampal activity seems to be 
phase-locked to the troughs of parietal and parahippocampal spindles. 
A characterization of these rhythms in terms of PRCs might shed light on the nature of this observation. 
Furthermore, knowing the response function of the system enables one to estimate cortical inputs based on the drift velocity of spiral waves \citep{biktasheva_2010}.

\begin{acknowledgements}
This work was supported by the Deut-sche Forschungsgemeinschaft (DFG)
within\\ 
SFB 654 ``Plasticity \& Sleep'' and the Graduate School for Computing in Medicine and Life Sciences 
funded by Germany�s Excellence Initiative [DFG GSC 235/1].
\end{acknowledgements}
\bibliographystyle{spbasic}
\bibliography{bib}   % name your BibTeX data base
% \begin{thebibliography}
% x
% \end{thebibliography}

% ------------
\begin{appendix}
\section{The network model} \label{sec:network_model}
In the original model by \citep{compte_cellular_2003} 1024 pyramidal neurons (see table~\ref{tab:Compte_model}) and 256 interneurons (see table~\ref{tab:IN}) are distributed equidistantly along a line of 5mm. 
The probability that two neurons, separated by a distance $x$, are connected is $P(x)=(\frac{1}{\sqrt{2\pi \sigma^2}})\exp(-x^2/2\sigma^2)$ with 
a synaptic footprint of $\sigma = 250\mathrm{\mu m}$ for excitatory connections and $\sigma=125\mathrm{\mu m}$ for inhibitory connections. 
The equations governing the synapses can be found in table~\ref{tab:Synapses}.
Each neuron makes $20 \pm 5$ connections to other neurons. In our simulations we used 256 pyramidal neurons and 64 interneurons. 
The network length and synaptic footprint was linearly scaled to preserve the properties of the original model. 
We applied periodic boundary conditions.
\begin{table*}
\caption{Regular-spiking pyramidal neurons}
\label{tab:Compte_model}       % Give a unique label
\begin{tabular}{lll}
\hline\noalign{\smallskip}
description & equations & parameters\\
  \noalign{\smallskip}\hline\noalign{\smallskip}
 somatic voltage 	& $C_\mathrm m A_\mathrm s \frac{\mathrm{d} V_\mathrm s}{\mathrm{d} t} = 
			  - A_\mathrm s (I_\mathrm L + I_\mathrm{Na} + I_\mathrm K 
			  + I_\mathrm A + I_\mathrm{KS} + I_\mathrm{KNa}) - $ 					& $C_\mathrm m=1\mu\mathrm{ F /cm^2}$ \\
			& $ - I_\mathrm{syn,s} - g_\mathrm{sd}(V_\mathrm s - V_\mathrm d) + I_\mathrm{ext} $ 	& $A_\mathrm s=0.015\mathrm{mm^2}$ \\
			&											& $g_\mathrm{sd} = (1.75 \pm 0.1)\mu\mathrm{S}$ \\
\noalign{\smallskip}\hline\noalign{\smallskip}
dendritic voltage  	& $ C_\mathrm m A_\mathrm d \frac{\mathrm{d} V_\mathrm d}{\mathrm{d} t}   
			  = - A_\mathrm d (I_\mathrm{Ca} + I_\mathrm{KCa} + I_\mathrm{NaP} + I_\mathrm{AR})- $ 	& $A_\mathrm d=0.035\mathrm{mm^2}$ \\
			& $ - I_\mathrm{syn,d} - g_\mathrm{sd}(V_\mathrm d - V_\mathrm s) + I_\mathrm{ext}$ 	& \\
			& $\frac{\mathrm{d} m}{\mathrm{d} t}  = \phi \left[\alpha_x(V)(1-m)-\beta_m(V)m \right]$& \\
			& $\frac{\mathrm{d} m}{\mathrm{d} t}  = \phi \left[m_\infty(V) - m \right]/ \tau_m (V)$ & \\
\noalign{\smallskip}\hline\noalign{\smallskip}
leakage current 	& $I_L = g_\mathrm{L} (V-V_\mathrm{L})$							& $V_\mathrm L = (-60.95 \pm 0.3)\mathrm{mV}$ \\
			& 											& $g_\mathrm L=(0.067 \pm 0.0067)\mathrm{mS/cm^2}$ \\
\noalign{\smallskip}\hline\noalign{\smallskip}
spiking  		& $I_\mathrm{Na}=g_\mathrm{Na} m_{\mathrm{Na},\infty}^3 h_\mathrm{Na}(V-V_\mathrm{Na})$		& $g_\mathrm{Na}= 50\mathrm{mS/cm^2}$ \\
sodium current  	& $m_{\mathrm{Na},\infty}=\alpha_{m_\mathrm{Na}}/(\alpha_{m_\mathrm{Na}} + \beta_{m_\mathrm{Na}})$ & $V_\mathrm{Na} = 55\mathrm{mV}$  \\
			& $\alpha_{m_\mathrm{Na}}=0.1(V + 33)/[1- \exp(-(V+33)/10]$				& $\phi=4$ \\
			& $\beta_{m_\mathrm{Na}}= 4 \exp(-(V + 53.7)/12)$ 					& \\
			& $\frac{\mathrm{d} h_{\mathrm{Na}}}{\mathrm{d} t}  = \phi \left[\alpha_{h_\mathrm{Na}}(V)
			  (1-h_\mathrm{Na})-\beta_{h_\mathrm{Na}}(V)h_\mathrm{Na} \right]$ & \\
			& $\alpha_{h_\mathrm{Na}} = 0.07 \exp(-(V + 50)/10)$ &  \\
			& $\beta_{h_\mathrm{Na}}= 1 /[1 + \exp(-(V + 20)/10)]$&  \\
\noalign{\smallskip}\hline\noalign{\smallskip}
spiking 		& $I_\mathrm{K} = g_\mathrm{K} h_\mathrm{K}^4(V-V_\mathrm{K})$ 				& $\phi=4$ \\
potassium current 	& $\frac{\mathrm{d} h_{\mathrm{K}}}{\mathrm{d} t}  = \phi \left[\alpha_{h_\mathrm{K}}(V)
			  (1-h_\mathrm{K})-\beta_{h_\mathrm{K}}(V)h_\mathrm{K} \right]$ 			& $g_\mathrm K = 10.5\mathrm{mS/cm^2}$ \\
			& $\alpha_{h_\mathrm{K}}=0.01(V + 34 )/[ 1 - \exp (-(V+34)/10)]$ 			& $V_\mathrm K =-100\mathrm{mV}$ \\
			& $\beta_{h_\mathrm{K}}=0.125 [ \exp (-(V + 44)/25)]$ &   \\
\noalign{\smallskip}\hline\noalign{\smallskip}
fast 			& $I_\mathrm A = g_\mathrm A m_{\mathrm{A},\infty} h_{\mathrm A} (V - V_\mathrm K)$ 	& $g_\mathrm A = 1\mathrm{mS/cm^2}$\\
inactivating 		& $m_{\mathrm{A},\infty} = 1/[1 + \exp(-(V +50 )/20)]$  				& $\tau _{h_\mathrm{A}}=15\mathrm{ms}$\\
current  		& $\frac{\mathrm{d} h_{\mathrm A}}{\mathrm{d} t}  = \left(h_{\mathrm A,\infty}(V) - h_{\mathrm A} \right)/ \tau_{h_\mathrm A}$ & \\
			& $h_{\mathrm A,\infty} = 1/[ \exp (- (V + 80)/6)]$ 					& \\
\noalign{\smallskip}\hline\noalign{\smallskip}
non-inactivating 	& $I_\mathrm{KS}=g_\mathrm{KS}m_\mathrm{KS}(V -V_\mathrm K)$ 				& $g_\mathrm{KS}=0.576\mathrm{mS/cm^2}$ \\
K$^+$--channel  	& $\frac{\mathrm{d} m_\mathrm{KS}}{\mathrm{d} t}  = \left(m_{\mathrm{KS},\infty}(V) - m_\mathrm{KS} \right)/ \tau_{m_\mathrm{KS}}$\\
			& $m_{\mathrm{KS},\infty}=1/[1 + \exp (-(V + 34.5)/6.5]$ 				&  \\
			& $\tau_{m_\mathrm{KS}} = 8/[\exp(-(V+55)/30) + \exp((V + 55)/30)]$ & \\ 
\noalign{\smallskip}\hline\noalign{\smallskip}
non-inactivating	& $I_\mathrm{NaP}=g_\mathrm{NaP}m^3_{\mathrm{NaP},\infty} (V - V_\mathrm{Na})$ 		& $g_\mathrm{NaP}=0.0686\mathrm{mS/cm^2}$\\ 
sodium channel 		& $m_{\mathrm{NaP},\infty}=1/[1 + \exp(-(V + 55.7)/7.7)]$ 				&  \\ 
\noalign{\smallskip}\hline\noalign{\smallskip}
hyperpolarization 	& $I_\mathrm{AR}=g_{\mathrm{AR}}h_{\mathrm{AR},\infty}(V-V_\mathrm K)$ 			& $g_\mathrm{AR} = 0.0257\mathrm{mS/cm^2}$\\
de-inactivated 	& $h_{\mathrm{AR},\infty}=1/[1 + \exp((V+75)/4]$ &  \\
channel 	 	& & \\
\noalign{\smallskip}\hline\noalign{\smallskip}
high-threshold  	& $I_\mathrm{Ca}=g_{\mathrm{Ca}}m_{\mathrm{Ca},\infty}^2(V - V_\mathrm{Ca})$ 		& $g_\mathrm{Ca} = 0.43\mathrm{mS/cm^2}$  \\
Ca$^{2+}$--channel 	& $m_{\mathrm{Ca},\infty}=1/[1 + \exp(-(V + 20)/9)]$ 					& $V_\mathrm{Ca}=120\mathrm{mV}$ \\
\noalign{\smallskip}\hline\noalign{\smallskip}
calcium dependent 	& $I_\mathrm{KCa}=g_\mathrm{KCa}[\mathrm{Ca}^{2+}]/([\mathrm{Ca}^{2+}] + K_D)(V - V_\mathrm K)$ & $g_\mathrm{KCa}=0.57\mathrm{mS/cm^2}$\\
potassium channel 	& $\mathrm d [\mathrm{Ca}^{2+}]/ \mathrm d t = - \alpha_\mathrm{Ca}A_dI_\mathrm{Ca} - [\mathrm{Ca}^{2+}]/\tau_\mathrm{Ca}$ & 
			  $\alpha_\mathrm{Ca}=0.005\mu\mathrm{M}/(\mathrm{nA}\cdot \mathrm{ms})$\\
			& & $\tau_\mathrm{Ca}=150\mathrm{ms}$ \\
\noalign{\smallskip}\hline\noalign{\smallskip}
sodium dependent 	& $I_\mathrm{KNa}= g_\mathrm{KNa}w_\infty([\mathrm{Na}^+])(V-V_\mathrm K)$ & $g_\mathrm{KNa}=1.33\mathrm{mS/cm^2}$ \\
potassium channel 	& $w_\infty= 0.37 / [1 + (38.7/[\mathrm{Na}^+])^{3.5}]$ &  \\
\noalign{\smallskip}\hline\noalign{\smallskip}
sodium dynamics 	& $\mathrm d [\mathrm{Na}^+]/\mathrm d t = -\alpha_\mathrm{Na}(A_\mathrm s I_\mathrm{Na} + A_\mathrm d I_\mathrm{NaP})$ & 
			  $\alpha_\mathrm{Na}=0.01\mathrm{mM}/(\mathrm{nA} \cdot \mathrm{ms})$\\ 
			& $-R_\mathrm{pump}\{[\mathrm{Na}^+]^3/([\mathrm{Na}^+]^3+15^3) - 
			  [\mathrm{Na}^+]_{\mathrm{eq}}^3/([\mathrm{Na}^+]_{\mathrm{eq}}^3 + 15^3)\}$ 		& 
			  $R_\mathrm{pump}=0.018\mathrm{mM/ms}$	\\
			& & $[\mathrm{Na}^+]_{\mathrm{eq}}=9.5\mathrm{mM}$ \\			
\noalign{\smallskip}\hline
\end{tabular}
\end{table*}

\begin{table*}
\caption{Fast-spiking inhibitory interneurons}
\label{tab:IN}       % Give a unique label
\begin{tabular}{lll}
\hline\noalign{\smallskip}
description & equations & parameters\\
\noalign{\smallskip}\hline\noalign{\smallskip}
somatic voltage  	& $C_\mathrm m A_\mathrm i \frac{\mathrm{d} V_\mathrm i}{\mathrm{d} t} = 
			-A_\mathrm i(I_\mathrm L + I_\mathrm{Na} + I_\mathrm K) - I_\mathrm{syn,i} + I_\mathrm{ext}$ & $A_\mathrm i= 0.02~\mathrm{mm}^2$\\
			& 											& $g_\mathrm{Na}=35~\mathrm{mS/cm^2}$ \\
\noalign{\smallskip}\hline\noalign{\smallskip}
leak current  		& $I_\mathrm L =g_\mathrm L (V-V_\mathrm L)$	& $g_\mathrm L=(0.1025 \pm 0.0025)\mathrm{mS/cm^2}$\\
			&						& $V_\mathrm L = (-63.8 \pm 0.15)\mathrm{mV}$\\
\noalign{\smallskip}\hline\noalign{\smallskip}
spiking		   	& $I_\mathrm{Na}=g_\mathrm{Na}m_{\mathrm{Na},\infty}h_\mathrm{Na}(V - V_\mathrm{Na})$ 	& $g_\mathrm {Na}=35~\mathrm{mS/cm^2}$ \\
sodium current		& $m_{\mathrm{Na},\infty}=\alpha_{m_\mathrm{Na}}/(\alpha_{m_\mathrm{Na}}+\beta_{m_\mathrm{Na}})$ & $V_\mathrm{Na}=55~\mathrm{mV}$ \\
			& $\alpha_{m_\mathrm{Na}}=0.5(V+35)/[1-\exp(-(V+35)/10)]$ 				& \\
			& $\beta_{m_\mathrm{Na}}=20 \exp(-(V + 60)/18)$						& \\
			& $\frac{\mathrm{d} h_{\mathrm{Na}}}{\mathrm{d} t}  = \alpha_{h_\mathrm{Na}}(V)
			  (1-h_\mathrm{Na})-\beta_{h_\mathrm{Na}}(V)h_\mathrm{Na}$				&\\
			& $\alpha_{h_\mathrm{Na}}=0.35 \exp(-(V+58)/20)$ 					& \\
			& $\beta_{h_\mathrm{Na}}=5/[1+\exp(-(V+28)/10)]$					& \\
\noalign{\smallskip}\hline\noalign{\smallskip}
slow 			& $I_\mathrm K = g_\mathrm K m_\mathrm K ^4 (V- V_k)$			& $g_\mathrm K=9~\mathrm{mS/cm^2}$\\
potassium current	& $\frac{\mathrm{d} m_{\mathrm{K}}}{\mathrm{d} t}  = \alpha_{m_\mathrm{K}}(V)
			  (1-m_\mathrm{K})-\beta_{m_\mathrm{K}}(V)m_\mathrm{K}$			&  $V_\mathrm K = -90\mathrm{mV}$\\
			& $\alpha_{m_\mathrm K} =0.05 (V + 34)/[1 - \exp (-(V + 34)/10)]$	& \\
			&$\beta_{m_\mathrm K} = 0.625 \exp(-(V + 44)/80)$ 			& \\
\noalign{\smallskip}\hline
\end{tabular}
\end{table*}
\begin{table*}
\caption{Synapses}
\label{tab:Synapses}       % Give a unique label
\begin{tabular}{lll}
\hline\noalign{\smallskip}
description & equations & parameters\\
\noalign{\smallskip}\hline\noalign{\smallskip}
AMPA synapses	& $I_\mathrm{syn}= g_\mathrm{syn}s(V-V_\mathrm{syn})$			& $\alpha = 3.48$ \\
		& $\frac{\mathrm ds}{\mathrm dt}=\alpha f(V_\mathrm{pre}) -s/\tau$	& $\tau = 2\mathrm{ms}$	\\
		& $f(V_\mathrm{pre}) = 1/\left[1+ \exp(-(V\mathrm{pre} -20)/2) \right]$ & $V_\mathrm{syn}=0\mathrm V$ \\ 
		&									&$g_\mathrm{EE}^\mathrm{AMPA}=5.4 \mathrm{nS}$\\
		&									&$g_\mathrm{EI}^\mathrm{AMPA}=2.25 \mathrm{nS}$\\
\noalign{\smallskip}\hline\noalign{\smallskip}
NMDA synapses	& $\frac{\mathrm ds}{\mathrm dt}=\alpha_s(1-s)x - s/\tau_\mathrm{s}$ 	& $\alpha_\mathrm{s} = 0.5$ \\
		& $\frac{\mathrm dx}{\mathrm dt}=\alpha_x f(V_\mathrm{pre})- x/\tau_\mathrm{x} $
											& $\tau_\mathrm{s} = 100\mathrm{ms}$\\
		& $f(V_\mathrm{pre}) = 1/\left[1+ \exp(-(V\mathrm{pre} -20)/2) \right]$ & $\alpha_\mathrm{x} = 3.48$\\		
		& 	&  $\tau_\mathrm{x} =2\mathrm{ms} $ \\
		& 	&  $V_\mathrm{syn} = 0\mathrm{mV}$\\
		&	&  $g_\mathrm{EE}^\mathrm{NMDA}=0.9 \mathrm{nS}$\\
		&	&  $g_\mathrm{EI}^\mathrm{NMDA}=0.5 \mathrm{nS}$\\
\noalign{\smallskip}\hline\noalign{\smallskip}
GABA synapses	& $I_\mathrm{syn}= g_\mathrm{syn}s(V-V_\mathrm{syn})$			& $\alpha = 1$ \\
		& $\frac{\mathrm ds}{\mathrm dt}=\alpha f(V_\mathrm{pre}) -s/\tau$	& $\tau = 10\mathrm{ms}$	\\
		& $f(V_\mathrm{pre}) = 1/\left[1+ \exp(-(V\mathrm{pre} -20)/2) \right]$ & $V_\mathrm{syn}=-70\mathrm{mV}$ \\ 
		&	&  $g_\mathrm{IE}=4.15 \mathrm{nS}$\\
\noalign{\smallskip}\hline
\end{tabular}
\end{table*}
\end{appendix}

% \begin{figure}
% % Use the relevant command to insert your figure file.
% % For example, with the graphicx package use
%   \includegraphics{example.eps}
% % figure caption is below the figure
% \caption{Please write your figure caption here}
% \label{fig:1}       % Give a unique label
% \end{figure}
% %
% % For two-column wide figures use
% \begin{figure*}
% % Use the relevant command to insert your figure file.
% % For example, with the graphicx package use
%   \includegraphics[width=0.75\textwidth]{example.eps}
% % figure caption is below the figure
% \caption{Please write your figure caption here}
% \label{fig:2}       % Give a unique label
% \end{figure*}
% \begin{table}
% % table caption is above the table
% \caption{Please write your table caption here}
% \label{tab:1}       % Give a unique label
% % For LaTeX tables use
% \begin{tabular}{lll}
% \hline\noalign{\smallskip}
% first & second & third  \\
% \noalign{\smallskip}\hline\noalign{\smallskip}
% number & number & number \\
% number & number & number \\
% \noalign{\smallskip}\hline
% \end{tabular}
% \end{table}



\end{document}